\documentclass[a4paper]{jpconf%
												}

\usepackage[ansinew]{inputenc}
\usepackage{graphicx,
						epstopdf,
						cite,
						SIunits,
						amsmath}

\begin{document}

\title{In-medium QCD sum rules for \(\boldsymbol D\) mesons:\\ A projection method for higher order contributions}

\author{T Buchheim, T Hilger and B K\"{a}mpfer}

\address{Helmholtz-Zentrum Dresden-Rossendorf, PF 510119, D-01314 Dresden, Germany}
\address{Technische Universit\"{a}t Dresden, Institut f\"{u}r Theoretische Physik, D-01062 Dresden, Germany}

\ead{t.buchheim@hzdr.de}

\begin{abstract}
\(D\) mesons serve as excellent probes of hot and/or dense strongly interacting matter. They can provide insight into the restoration of chiral symmetry. The chiral condensate as well as other chirally odd condensates, such as certain four-quark condensates, are linked to order parameters of spontaneous chiral symmetry breaking. Thus, the evaluation of these higher order condensate contributions in the framework of QCD sum rules is of high interest. We present a general method for projecting Lorentz indices of ground state expectation values providing a crucial step towards a comprehensive calculation of higher order corrections to the operator product expansion of hadrons, especially \(D\) mesons, in a strongly interacting medium.
\end{abstract}

\section{Introduction}

Among the central issues of hadron physics is chiral symmetry, its breaking in vacuum and its restoration in medium. Chiral symmetry is explicitly broken due to non-zero quark masses leaving the mass term of the Lagrangian not invariant under chiral transformations. If only light quark flavors (up, down) are under consideration the assumption of massless quarks is justified. However, chiral symmetry is broken spontaneously, because the QCD ground state is not invariant under chiral transformations.

The chiral condensate \(\langle \bar{q}q \rangle\) is linked to order parameters of the spontaneous chiral symmetry breaking. This chirally odd condensate, i.e.\ a condensate which is not invariant under chiral transformations, is non-zero and thus, signals spontaneous symmetry breaking, similarly to the vector and axial-vector mixing under axial transformations. Besides the chiral condensates also further condensates are connected to order parameters of spontaneous chiral symmetry breaking, e.g.\ chirally odd four-quark-condensates. Chiral spontaneous symmetry breaking gets apparent due to the mass splitting of chiral partner mesons. For instance, in the light-quark sector, the axial vector meson \(a_1\) 
is significantly heavier than its chiral partner, the vector meson \(\rho\)
\cite{pdg}.

In a strongly interacting medium the situation changes drastically: We have to deal with non-zero temperatures \(T\) and baryon densities \(n\). In leading-order, non-zero temperatures are modeled by a pion gas and finite densities by ambient nucleons. The chiral condensate changes in a medium according to (cf.~\cite{hatsukoike,cohen95})
\begin{align}
	\langle \bar{q}q \rangle_{T,n} = \langle \bar{q}q \rangle \left( 1 - \frac{T^2}{8f^2_\pi} - \frac{\sigma_N\, n}{m_\pi^2 f_\pi^2} \right) \, ,
\end{align}
where \(f_\pi\) is the pion decay constant, \(m_\pi\) the pion mass and \(\sigma_N\) is the nucleon sigma term. Obviously the numerical value of the chiral condensate decreases for increasing temperature and density; chiral restoration would be accompanied by \(\langle \bar{q}q \rangle_{T,n} = 0 \).

The impact of further chirally odd condensates, e.g.\ certain four-quark condensates behaving as the chiral condensate, is therefore of utmost interest in relation to chiral restoration. First investigations have been performed in the light-quark meson sector, e.g.\ a significant impact of chirally odd four-quark condensates was found in the framework of rho meson sum rules \cite{hilger12}. We aim at extending these evaluations to \(Qq\) meson systems, containing an heavy (\(Q\)) and a light (\(q\)) valence quark.

Choosing certain chiral transformations restricted to light quarks, one can derive an invariant Lagrangian \cite{hilger11}. However, vector and axial-vector currents as well as scalar and pseudo-scalar currents mix under these transformations
, i.e.\ also in the \(Qq\) sector certain chirally odd condensates determine the mass splitting of chiral partner mesons. Again, chiral symmetry restoration in medium is supposed to be accompanied by the vanishing of chirally odd condensates.

Theoretical investigations of four-quark condensates of \(Qq\) meson systems in medium is of large interest, since the envisaged experiments of the CBM and Panda collaborations at FAIR \cite{CBM,Panda} include the analysis of medium modifications of \(D\) mesons.

\section{In-medium QCD sum rules of \(\boldsymbol D\) mesons: operator product expansion}

QCD sum rules proved to be a successful tool to extract spectral properties of hadrons in vacuum from the theory of strong interaction \cite{svz79,rry}. During the last two decennia an extension to hadrons in an strongly interacting environment has given insight into their medium modifications \cite{hatsukoike,cohen95}, which are of large contemporary interest. The sum rule method utilizes the causal current-current correlator
\begin{align}
	\Pi (q) = i \int d^4x\; e^{iqx} \langle \mathrm{T}[J(x) J^\dagger (0)] \rangle \, ,
	\label{cccorr}
\end{align}
where \(\mathrm{T}[\ldots]\) means time ordering and \(J\) denotes a current reflecting the quark content and quantum numbers of the hadron under consideration. The correlator (\ref{cccorr}) can be related to the hadrons spectral function via a dispersion relation which provides a link of QCD, formulated in quark degrees of freedom, and hadronic phenomenology. The evaluation of \(\Pi(q)\) employs Wilson's operator product expansion (OPE) \cite{wilson}.

The OPE is an expansion at operator level. The operator product is expanded in an asymptotic series of local operators \(\mathcal{O}_i\) with increasing mass dimension:
\begin{align}
	J(x) J^\dagger (0) = \sum_i C_i(x) \mathcal O_i(0) \, .
\end{align}
The coefficients \(C_i\) in this series expansion are the Wilson coefficients. QCD provides three operators as basic elements for this expansion: quark operators \(q\), the gluonic field strength tensor \(G_{\mu\nu}\) and the covariant derivative \(D_\mu\), which appear in infinitely combinations building the \(\mathcal{O}_i\). We consider the expectation value of the current-operator product, thus, we need to use the expectation values of the operators of the expansion. These expectations values are the condensates, characterizing the complex QCD ground state. In vacuum they have certain numerical values, which may be universally used for sum rule evaluation.

QCD cannot be treated perturbatively for low momentum transfers and long distance phenomena due to the running strong coupling \(\alpha_s\). One of the beneficial features of the OPE is the separation of scales \cite{shuryakbook}, i.e.\ long range effects are absorbed into the condensates and the Wilson coefficients can be calculated by means of perturbation theory. In the light-quark sector, all vacuum condensates up to mass dimension 6 have been identified and their Wilson coefficients have been determined, unlike the in-medium OPE for \(D\) mesons. Their coefficients are determined up to mass dimension 5 for in-medium situations \cite{hilger09}. The complete in-medium contributions of four-quark condensates are determined for light mesons only \cite{thomas05}. The three-gluon condensate contribution, also of mass dimension 6, has been calculated for vacuum situations so far \cite{nikrad83}. This work addresses crucial computational steps towards a comprehensive calculation of Wilson coefficients of higher order condensates for in-medium situations, especially  for four-quark condensates of \(Qq\) mesons. New medium specific condensates come into play, which contain heavy quark operators.

\section{Four-quark condensates}

Only expectation values of operators which carry the assumed symmetries of the QCD ground state are considered condensates \cite{jin}. Formally, condensates are expectation values of hermitian operator products, which are to be Lorentz and Dirac scalars as well as color singlets. They are invariant under parity and time reversal transformations. Thus, the projection of Dirac, color and Lorentz indices of the expectation values has to be performed. The projection of Dirac and color indices relies on orthogonal bases provided by the Clifford algebra and the generators and unity element of the color group, respectively \cite{thomas07}. Projection of Lorentz indices is more involved, in particular for in-medium situations.

\begin{figure}%
\centering
\includegraphics[width=3cm]{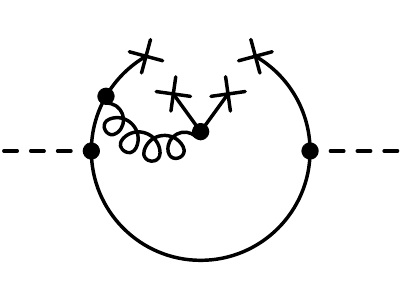}%
\hfill
\includegraphics[width=3cm]{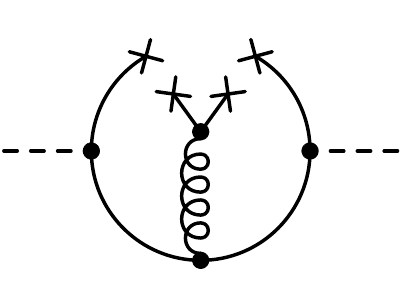}%
\hfill
\includegraphics[width=3cm]{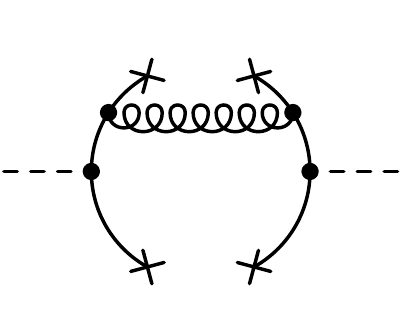}%
\hfill
\includegraphics[width=3cm]{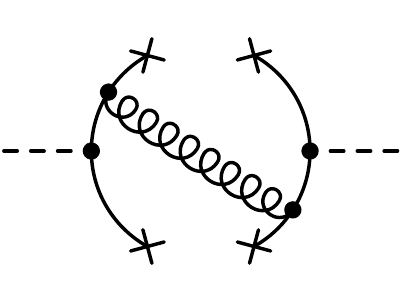}%
\caption{Diagrammatic representation of four-quark condensate contributions.}%
\label{diagrams}%
\end{figure}

In leading order, four-quark condensates contribute to \(D\) meson sum rules on tree-level in order \(\alpha_s\). The four-quark condensate contributions in a diagrammatic representation (cf.\ figure~\ref{diagrams}) are of two kinds: the meson momentum either flows through a quark or a gluon line \cite{svz79}. The depicted diagrams yield condensate contributions proportional to \(\langle \bar{q}_i \Gamma D_\mu D_\nu D_\lambda q_i \rangle\), \(\langle \bar{q}_i \Gamma D_\mu G_{\nu\lambda} q_i \rangle\) and \(\langle \bar{q} t^A \Gamma_1 q \bar{Q} t^A \Gamma_2 Q \rangle\) where \(q_i\) symbolizes either \(Q\) or \(q\), \(t^A\) stands for a generator or the unity element of the color group and \(\Gamma\), \(\Gamma_i\) denote basis elements of the Clifford algebra, carrying themselves up to two Lorenz indices each. The projection of these uncontracted Lorentz indices leads to 24 condensates in the \(Qq\) sector, which are invariant under parity and time reversal transformations.

\section{Projection of Lorentz Indices}

In contrast to Dirac and color projections, which rely on an expansion in an orthogonal basis, projecting Lorentz indices is based on given elements which carry Lorentz structures. In vacuum, these are the metric tensor \(g_{\mu\nu}\) and the totally anti-symmetric pseudo-tensor \(\varepsilon_{\mu\nu\lambda\sigma}\). For investigations of the temperature and/or density dependence of condensates, a further Lorentz structure is available for projection. Due to the presence of the ambient medium the Poincar\'{e} invariance is broken, and the medium velocity \(v_\mu\) needs to be considered additionally for projections. It has wide ranging implications, as it allows for the projection of expectation values with an odd number of Lorentz indices which results in a multitude of new condensates.

The ansatz for an expectation value \( \langle O_{\vec{\mu}_n} \rangle \equiv \langle O_{\mu_1 \ldots \mu_n} \rangle \) carrying \(n\) Lorentz indices reads
\begin{align}
	\langle O_{\vec{\mu}_n} \rangle = \mathbf{a} \cdot \mathbf{p}_{\vec{\mu}_n} \, ,
	\label{eq:LProjAnsatz}
\end{align}
where \( \mathbf{p}_{\vec{\mu}_n} \equiv \mathbf{p}_{\mu_1 \ldots \mu_n} \) denotes the vector of all projection structures and \(\mathbf{a}\) is the vector of the corresponding coefficients. Contracting Eq.~(\ref{eq:LProjAnsatz}) with \( \mathbf{p}^{\vec{\mu}_n} \) yields
\begin{align}
	&   \langle O_{\vec{\mu}_n} \rangle \mathbf{p}^{\vec{\mu}_n} =  \mathbf{p}^{\vec{\mu}_n} \left[ \mathbf{p}_{\vec{\mu}_n} \cdot \mathbf{a} \right] = \left[ \mathbf{p}^{\vec{\mu}_n} ( \mathbf{p}_{\vec{\mu}_n})^\text{T} \right] \mathbf{a} = P_n \mathbf{a} \, ,
\end{align}
where we have defined the matrix \(P_n=\mathbf{p}^{\vec{\mu}_n} ( \mathbf{p}_{\vec{\mu}_n} )^\text{T}\). Defining a vector \( \mathbf{c} = \langle O_{\vec{\mu}_n}\rangle \mathbf{p}^{\vec{\mu}_n} \), which contains the expectation values contracted with all projection structures, we obtain
\begin{align}
	\mathbf{a} & = P_n^{-1} \mathbf{c} \, .
\end{align}
Thus, the projection formula for \( n \) Lorentz indices reads in compact form
\begin{align}
	\langle O_{\vec{\mu}_n} \rangle & = \left( P_n^{-1} \mathbf{c} \right) \cdot \mathbf{p}_{\vec{\mu}_n}
	\label{eq:LProjRes1}
\end{align}
or more explicitly
\begin{align}
	\langle O_{\vec{\mu}_n} \rangle &	= \left\{ \left[\mathbf{p}^{\vec{\nu}_n} ( \mathbf{p}_{\vec{\nu}_n} )^\text{T}\right]^{-1} \langle O_{\vec{\kappa}_n} \rangle \mathbf{p}^{\vec{\kappa}_n} \right\} \cdot \mathbf{p}_{\vec{\mu}_n} \, .
\label{eq:LProjRes2}
\end{align}
The general projection vectors \(\mathbf{p}_{\vec{\mu}_n}\) for \(n\leq5\) have the following structures:
\begin{subequations}
\label{eq:genLProjVec}
	\begin{align}
 		\mathbf{p}_{\mu} = &\; v_\mu \, , \\[2mm]
		\mathbf{p}_{\mu\nu} = &\; (g_{\mu\nu}, \nonumber\\[1mm]
		&\;	v_\mu v_\nu)^\mathrm{T} \, , \\[2mm]
		\mathbf{p}_{\mu\nu\lambda} = &\; ( \varepsilon_{\mu\nu\lambda\alpha} v^\alpha, \nonumber\\[1mm]
		&\;	v_\mu g_{\nu\lambda},\, v_\nu g_{\mu\lambda},\, v_\lambda g_{\mu\nu}, \nonumber\\[1mm]
		&\;	v_\mu v_\nu v_\lambda )^\mathrm{T} \, , \\[2mm]
		\mathbf{p}_{\mu\nu\lambda\sigma} = &\; ( \varepsilon_{\mu\nu\lambda\sigma}, \nonumber\\[1mm]
		&\; v_\mu \varepsilon_{\nu\lambda\sigma\alpha} v^\alpha , v_\nu \varepsilon_{\mu\lambda\sigma\alpha} v^\alpha, v_\lambda \varepsilon_{\mu\nu\sigma\alpha} v^\alpha,   \nonumber\\[1mm]
		&\;	g_{\mu\nu}g_{\lambda\sigma},\, g_{\mu\lambda}g_{\nu\sigma},\, g_{\mu\sigma}g_{\nu\lambda}, \nonumber\\[1mm]
		&\;	v_\mu v_\nu g_{\lambda\sigma},\, v_\mu v_\lambda g_{\nu\sigma},\, v_\mu v_\sigma g_{\nu\lambda},\, v_\lambda v_\sigma g_{\mu\nu},\, v_\nu v_\sigma g_{\mu\lambda},\, v_\nu v_\lambda g_{\mu\sigma}, \nonumber\\[1mm]
		&\;	v_\mu v_\nu v_\lambda v_\sigma)^\mathrm{T} \, , \\[2mm]
		\mathbf{p}_{\mu\nu\lambda\sigma\rho} = &\; ( v_\mu \varepsilon_{\nu\lambda\sigma\rho},\, v_\nu \varepsilon_{\mu\lambda\sigma\rho},\, v_\lambda \varepsilon_{\mu\nu\sigma\rho},\, v_\sigma \varepsilon_{\mu\nu\lambda\rho}, \nonumber\\[1mm]
		&\; g_{\mu\nu} \varepsilon_{\lambda\sigma\rho\alpha} v^\alpha, g_{\mu\lambda} \varepsilon_{\nu\sigma\rho\alpha} v^\alpha, g_{\mu\sigma} \varepsilon_{\nu\lambda\rho\alpha} v^\alpha, g_{\nu\lambda} \varepsilon_{\mu\sigma\rho\alpha} v^\alpha, g_{\nu\sigma} \varepsilon_{\mu\lambda\rho\alpha} v^\alpha, g_{\lambda\sigma} \varepsilon_{\mu\nu\rho\alpha} v^\alpha, \nonumber\\[1.5mm]
		&\; v_\mu v_\nu \varepsilon_{\lambda\sigma\rho\alpha} v^\alpha, v_\mu v_\lambda \varepsilon_{\nu\sigma\rho\alpha} v^\alpha, v_\mu v_\sigma \varepsilon_{\nu\lambda\rho\alpha} v^\alpha, v_\nu v_\lambda \varepsilon_{\mu\sigma\rho\alpha} v^\alpha, v_\nu v_\sigma \varepsilon_{\mu\lambda\rho\alpha} v^\alpha, \nonumber\\
		&\; v_\lambda v_\sigma \varepsilon_{\mu\nu\rho\alpha} v^\alpha, \nonumber\\[1mm]
		&\; v_\mu g_{\nu\lambda}g_{\sigma\rho},\, v_\mu g_{\nu\sigma}g_{\lambda\rho},\, v_\mu g_{\nu\rho}g_{\lambda\sigma},\, v_\nu g_{\mu\lambda}g_{\sigma\rho},\, v_\nu g_{\mu\sigma}g_{\lambda\rho},\, v_\nu g_{\mu\rho}g_{\lambda\sigma},\, v_\lambda g_{\mu\nu}g_{\sigma\rho}, \nonumber\\
		&\; v_\lambda g_{\mu\sigma}g_{\nu\rho},\, v_\lambda g_{\mu\rho}g_{\nu\sigma},\, v_\sigma g_{\mu\nu}g_{\lambda\rho},\, v_\sigma g_{\mu\lambda}g_{\nu\rho},\, v_\sigma g_{\mu\rho}g_{\nu\lambda},\, v_\rho g_{\mu\nu}g_{\lambda\sigma},\, v_\rho g_{\mu\lambda}g_{\nu\sigma}, \nonumber\\
		&\; v_\rho g_{\mu\sigma}g_{\nu\lambda}, \nonumber\\[1mm]
		&\; v_\mu v_\nu v_\lambda g_{\sigma\rho},\, v_\mu v_\nu v_\sigma g_{\lambda\rho},\, v_\mu v_\nu v_\rho g_{\lambda\sigma},\, v_\mu v_\sigma v_\rho g_{\nu\lambda},\, v_\mu v_\lambda v_\rho g_{\nu\sigma}, \nonumber\\
		&\; v_\mu v_\lambda v_\sigma g_{\nu\rho},\, v_\nu v_\sigma v_\rho g_{\mu\lambda},\, v_\nu v_\lambda v_\rho g_{\mu\sigma},\, v_\nu v_\lambda v_\sigma g_{\mu\rho},\, v_\lambda v_\sigma v_\rho g_{\mu\nu}, \nonumber\\[1mm]
		&\; v_\mu v_\nu v_\lambda v_\sigma v_\rho)^\mathrm{T} \, ,
	\end{align}
\end{subequations}
where \(\mathbf{p}_{\vec{\mu}_n}\) is listed in the following way: If possible, projection structures incorporating the totally anti-symmetric pseudo-tensor are noted first, the following lines (blocks) contain projection structures incorporating the metric tensor with increasing number of quantities \(v_\mu\) (from line (block) to line (block)).

In order to separate vacuum and medium specific structures the projection is decomposed as
\begin{align}
	\langle O_{\vec{\mu}_n} \rangle = \langle O_{\vec{\mu}_n} \rangle^\text{vac} + \langle O_{\vec{\mu}_n} \rangle^\text{med} \, .
	\label{eq:LProjDecomp}
\end{align}
If the expectation value carries an odd number of Lorentz indices it is a purely medium specific condensate and a decomposition is obsolete. When dealing with an even number of Lorentz indices the vacuum projection \(\langle O_{\vec{\mu}_n} \rangle^\text{vac}\) is obtained by applying the depicted method, but using \(\mathbf{p}^\text{vac}_{\vec{\mu}_n}\) instead of \(\mathbf{p}_{\vec{\mu}_n}\) containing the vacuum specific elements \(g_{\mu\nu}\) and \(\varepsilon_{\mu\nu\lambda\sigma}\) only. The medium projection is obtained according to \( \langle O_{\vec{\mu}_n} \rangle^\text{med} = \langle O_{\vec{\mu}_n} \rangle - \langle O_{\vec{\mu}_n} \rangle^\text{vac} \).

\section{Summary}

Four-quark condensate contributions to the OPE of \(D\) mesons for general in-medium situations extend existing QCD sum rule evaluations. One of the major computational tasks for calculating contributions of higher order condensates, especially four-quark condensates, is the projection of Dirac, color and Lorentz indices. Generalization of the projection of Dirac and color indices is straight forward because they rely on an orthogonal basis and with regard to four-quark condensates one observes a close connection to Fierz transformations (cf.\ \cite{drukarev04,zhangbrauner}). However, the projection of Lorentz indices for in-medium situations is more involved. In this paper we present a general procedure which goes beyond the common itemization of explicit examples, e.g.\ given in \cite{jin,nikrad83}. 
Our method is now being implemented in the evaluation of Wilson coefficients of those four-quark condensates which appear as \(\alpha_s\) contributions containing terms intimately related to chiral symmetry.

\ack

This work is supported by BMBF 05P12CRGHE.

\section*{References}


\begin{thebibliography}{10}
\expandafter\ifx\csname url\endcsname\relax
  \def\url#1{{\tt #1}}\fi
\expandafter\ifx\csname urlprefix\endcsname\relax\def\urlprefix{URL }\fi
\providecommand{\eprint}[2][]{\url{#2}}

\bibitem{pdg}
Beringer J {\em et~al.\/} (Particle Data Group) 2012 {\em Phys. Rev. {\rm D}\/}
  {\bf 86} 010001

\bibitem{hatsukoike}
Hatsuda T, Koike Y and Lee S~H 1993 {\em Nucl. Phys. {\rm B}\/} {\bf 394} 221

\bibitem{cohen95}
Cohen T~D, Furnstahl R~J, Griegel D~K and Jin X 1995 {\em Prog. Part. Nucl.
  Phys.\/} {\bf 35} 221

\bibitem{hilger12}
Hilger T, Thomas R, K\"{a}mpfer B and Leupold S 2012 {\em Phys. Lett. {\rm B}\/}
  {\bf 709} 200

\bibitem{hilger11}
Hilger T, K\"{a}mpfer B and Leupold S 2011 {\em Phys. Rev. {\rm C}\/} {\bf 84}
  045202

\bibitem{CBM}
{CBM Collaboration} 2013
  {http://www.fair-center.eu/en/for-users/experiments/cbm.html}

\bibitem{Panda}
{PANDA Collaboration} 2013 {http://www-panda.gsi.de/framework/physics.php}

\bibitem{svz79}
Shifman M~A, Vainshtein A~I and Zakharov V~I 1979 {\em Nucl. Phys. {\rm B}\/}
  {\bf 147} 385

\bibitem{rry}
Reinders L~J, Rubinstein H and Yazaki S 1985 {\em Phys. Rept.\/} {\bf 127} 1

\bibitem{wilson}
Wilson K~G 1969 {\em Phys. Rev.\/} {\bf 179} 1499

\bibitem{shuryakbook}
Shuryak E~V 2004 {\em {The QCD vacuum, hadrons and superdense matter}\/} 2nd ed
  {Lecture Notes in Physics} ({World Scientific})

\bibitem{hilger09}
Hilger T, Thomas R and K\"{a}mpfer B 2009 {\em Phys. Rev. {\rm C}\/} {\bf 97}
  025202

\bibitem{thomas05}
Thomas R, Zschocke S and K\"{a}mpfer B 2005 {\em Phys. Rev. Lett.\/} {\bf 95}
  232301

\bibitem{nikrad83}
Nikolaev S~N and Radyushkin A~V 1983 {\em Nucl. Phys. {\rm B}\/} {\bf 213} 285

\bibitem{jin}
Jin X, Cohen T~D, Furnstahl R~J and Griegel D~K 1993 {\em Phys. Rev. {\rm C}\/}
  {\bf 47} 2882

\bibitem{thomas07}
Thomas R, Hilger T and K\"{a}mpfer B 2007 {\em Nucl. Phys. {\rm A}\/} {\bf 795} 19

\bibitem{drukarev04}
Drukarev E~G, Ryskin M~G, Sadovnikova V~A, Gutsche T and Faessler A 2004 {\em
  Phys. Rev. {\rm C}\/} {\bf 69} 065210

\bibitem{zhangbrauner}
Zhang T, Brauner T and Rischke D~H 2010 {\em JHEP\/} {\bf 1006} 064

\end{thebibliography}

\providecommand{\newblock}{}

\end{document}